\documentclass[namedreferences]{solarphysics}

\usepackage[hyperref,optionalrh,natbib]{spr-sola-addons} 
\usepackage{graphicx}        
\usepackage{epstopdf}
\usepackage{amssymb}        
\usepackage{color}           

\newcommand{\arcsec}{{$^{\prime\prime}$}}

\begin{document}
\begin{article}

\begin{opening}

\title{Oscillations in the 45\,--\,5000\,MHz Radio Spectrum\\ of the 18 April 2014 Flare}

\author[ addressref={aff1}]{\inits{M.}\fnm{Marian}~\lnm{Karlick\'{y}}}
\author[ addressref={aff2}, corref, email={rybak@astro.sk}]{\inits{J.}\fnm{J\'{a}n}~\lnm{Ryb\'{a}k}\orcid{0000-0003-3128-8396}}
\author[ addressref={aff3} ]{\inits{C.}\fnm{Christian}~\lnm{Monstein}}

\runningauthor{M. Karlick\'y, J. Ryb\'ak, and C. Monstein}
\runningtitle{Oscillations in Radio Spectrum of the 18 April 2014 Flare}

\address[id=aff1]{Astronomical Institute, Academy of Sciences of
                  the Czech Republic, 251 65 Ond\v{r}ejov, Czech Republic}
\address[id=aff2]{Astronomical Institute, Slovak Academy of Sciences,
                  Tatransk\'{a} Lomnica, Slovakia}
\address[id=aff3]{Institute for Astronomy, ETH Zurich, CH-8093 Zurich, 
                  Switzerland}

\begin{abstract}
Using a new type of oscillation map, made from the radio spectra by the
wavelet technique, we study the 18 April 2014 M7.3 flare
(SOL2014-04-18T13:03:00L245C017). 
We find a quasi--periodic character of this flare with periods in the range 
65\,--\,115\,seconds. At the very beginning of this flare, 
in connection with the drifting pulsation structure (plasmoid
ejection) we find the 65\,--\,115\,s oscillation phase drifting slowly towards
lower frequencies, which indicates an upward propagating wave initiated  at
the start of the magnetic reconnection. In the drifting pulsation structure
many periods (1\,--\,200 seconds) are found documenting multi--scale and
multi--periodic processes. On this drifting structure  fiber bursts with a
characteristic period of about one second are superimposed, whose
frequency drift is similar to that of the drifting 65\,--\,115\,s oscillation
phase. We also check periods found in this flare by 
{\it EUV Imaging Spectrometer} (EIS)/{\it Hinode} and 
{\it Interface Region Imaging Spectrograph} (IRIS) observations. 
We recognize the type III bursts (electron beams) as proposed,
but their time coincidence with the EIS and IRIS peaks is not very good. 
This is probably due to the radio spectrum beeing a whole--disk record 
consisting of all bursts from any location while 
the EIS and IRIS peaks are emitted only from locations of slits in the EIS 
and IRIS observations.
\end{abstract}
\keywords{Sun: flares --- Sun: radio radiation --- Sun: oscillations}

\end{opening}

\section{Introduction}

In X-ray, ultraviolet and radio emissions of solar flares, oscillations
and waves are commonly observed
~(\opencite{1984ApJ...279..857R,2003SoPh..218..183F,2005A&A...435..753W,2006A&A...452..343N,2010PPCF...52l4009N,2016ApJ...822....7K}).

These oscillations have periods ranging from sub--seconds to tens of
minutes
~(\opencite{2006A&A...460..865M,2008SoPh..253..117T,2010SoPh..261..281K,2010SoPh..267..329K,
2011A&A...525A..88M,2014ApJ...791...44H,2014A&A...569A..12N}).

For overviews of oscillations and waves, and their models, see the
papers by~\cite{2009SSRv..149..119N},~\cite{2016SoPh..291.3143V}, and
~\cite{2016SSRv..200...75N}.

In several papers various types of oscillations and waves were already
numerically studied: hot coronal loop oscillations \citep{2002ApJ...580L..85O},
slow MHD waves in solar coronal magnetic fields \citep{2003A&A...408..755D},
acoustic oscillations in solar and stellar flaring loops
\citep{2004A&A...414L..25N}, short quasi-periodic MHD waves in coronal
structures \citep{2005SSRv..121..115N}, slow magnetosonic standing waves in a
solar coronal loop \citep{2005A&A...436..701S}, impulsively generated slow
acoustic waves in a solar coronal loop \citep{2009EPJD...54..305J}, impulsively
generated wave trains in coronal loops \citep{2010ITPS...38.2243J}, slow
magnetoacoustic standing modes in a gravitationally stratified solar coronal
arcade \citep{2010A&A...521A..34K}, slow magnetosonic waves in active region
loops \citep{2012ApJ...754..111O}, standing kink modes in coronal loops
\citep{2014ApJ...784..101P}, and magnetoacoustic waves propagating along a
dense slab and Harris current sheet \citep{2014ApJ...788...44M}.

Recently, searching for signatures of the fast magnetosonic waves in the 1
August 2010 event, we developed a new type of map of oscillations based on the
wavelet transform analysis of radio spectra (\opencite{2017SoPh..292....1K}).

In the present paper, we construct such maps of oscillations for the 18 April
2014 flare that looks to be rich in various oscillations. Namely, 
oscillations were recognized in the global X--ray flux detected by the
{\it Fermi}/{\it Gamma-ray Burst Monitor} (GBM) instrument as well as in 
the spatially localized UV measurements of
the {\it Interface Region Imaging Spectrograph} (IRIS) and
{\it EUV Imaging Spectrometer} (EIS)/{\it Hinode} 
instruments (\opencite{2015ApJ...810...45B},
\opencite{2016ApJ...830..101B} and \opencite{2015ApJ...810....4B}).
Moreover as shown by \cite{2016ApJ...830..101B} in the interval of
these oscillations the profiles of the O{\sc\,IV}\,--\,Fe{\sc\,XVI} 
lines were redshifted and
the electron densities in the regions of the Fe XIV and Mg VII line generation 
were 4.6 $\times$ 10$^{10}$ cm$^3$ and 7.8 $\times$ 10$^9$ cm$^3$, respectively.
Further quasi--periodic oscillations in both position and Doppler velocities
were found in small--scale substructure within the flare ribbon
\citep{2015ApJ...810....4B}.

Therefore, in the paper, we not only want to know what oscillations are present
in radio emission of this flare, but we want to compare them with the
oscillations detected in X--rays and in the IRIS and EIS UV observations. Of
course, we also want to confirm or reject the beam origin idea of EIS
oscillation proposed by \cite{2016ApJ...830..101B}.

Because we analyze the radio spectra in decimetric and metric ranges, where
most of bursts are generated by the plasma emission mechanism (the emission
frequency depends on the plasma density in the radio source), the frequency of
oscillations on the map gives us an information about the altitude of these
oscillations in the solar atmosphere. Furthermore, a possible frequency drift
of oscillation phases indicates propagating waves.

The paper is structured as follows: In Section 2 we present the data and
methods of analysis. The results and their interpretations are
summarized in Section 3. Conclusions are in Section 4.

\section{Data and Methods of Their Analysis}

The radio spectrum was aquired during the 18 April 2014 flare ({\it
Hinode} Flare Catalog - event number 103690~(\opencite{2012SoPh..279..317W})).
This flare, classified as a GOES M7.3 flare, started at 12:31\,UT, peaked at
13:03\,UT, and ended at 13:20\,UT in NOAA active region 12036 (S17W40).
For more details about radio observations of this flare, see the
recent paper by \citet{2016ApJ...833...87C}.

The overall radio spectrum selected for our analysis consists of two
e--Callisto (International Network of Solar Radio Spectrometers)
radiospectrographs, and two Ond\v{r}ejov radiospectrographs for the time
interval 12:40\,--\,13:10\,UT. 
The details of the observations are as follows:

\begin{itemize}
\item The \href{http://soleil.i4ds.ch/solarradio/}{{\it Bleien BLEN7M radiospectrograph}} is
located at Bleien, Switzerland. Its frequency range, mean frequency resolution, and temporal
resolution are 175\,--\,870\,MHz, 3.61\,MHz, and 0.25 second respectively.

\item The \href{http://www.astro.gla.ac.uk/observatory/srt}{{\it Glasgow Small Radio Telescope}} is
located at Acre Road Observatory, Glasgow, Scotland. Its frequency range, mean frequency
resolution, and temporal resolution are 81\,--\,45\,MHz, 0.2\,MHz, and 0.25 second, respectively.

\item The \href{http://www.asu.cas.cz/~radio/}{{\it Ond\v{r}ejov radiospectrograph RT5}} is
located at Ond\v{r}ejov, Czech Republic. Radiospectrograph frequency range,
frequency resolution, and temporal resolution are 800\,--\,2000\,MHz, 4.7\,MHz,
and 10\,ms, respectively \citep{2008SoPh..253...95J}.

\item The \href{http://www.asu.cas.cz/~radio/}{{\it Ond\v{r}ejov radiospectrograph RT4}} is
located at Ond\v{r}ejov, Czech Republic. Radiospectrograph frequency range,
frequency resolution, and temporal resolution are 2000\,--\,5000\,MHz,
11.6\,MHz, and 1\,second, respectively \citep{1993SoPh..147..203J}.
\end{itemize}

During data selection all records available in the e--Callisto
archive{\footnote{see \tt
http://soleil80.cs.technik.fhnw.ch/solarradio/data/2002-20yy\_Callisto/ .}} for
this particular time interval are checked. Only radiospectrograms of the best
quality in the frequency ranges of interest are selected. Those data are
adapted using the dedicated software from the {\tt ethz} library in SolarSoft
\citep{1998SoPh..182..497F} according the dedicated manual\footnote{see {\tt
http://www.e-callisto.org/Software/phoenix\_howto.html}.}. The e-Callisto data
are not calibrated in intensity and these data are a logarithmic representation of
solar radio burst activity \citep{2005SoPh..226..143B}.

All radio data are re--sampled to the final temporal sampling of 1 second except a
particular additional data set of the Ond\v{r}ejov 800\,--\,2000\,MHz data in
the time interval 12:47\,--\,12:49\,UT, for which the 0.1\,second sampling is used.

At some frequencies, artificial radio noise heavily affecs the acquired signal.
Such parts of the data are excluded and they are marked by black bands
in all plots of the radiospectrograms.
The general radio spectrum composed from the individual radio spectra of all four
radiospectrographs is shown in the left column of Figure\,\ref{figure3}.
\\
\\
For analysis of the radio spectra we utilize a novel method introduced in
our recent article \citep{2017SoPh..292....1K}. 
The wavelet transform (WT) is a core of this method providing a clear detection 
of time--frequency evolution of the significantly strong radio signal 
wave--patterns.
This method detects the identified wave--pattern signal power which is
located in the selected period range at any temporal moment and radio frequency.

Due to possible multiplicity of the non--stationary periodicities
the investigated periodicities are limited to maximum power in a
selected narrower WT period range of particular interest.
The WT computational algorithm of \cite{1998BAMS...79...61T} is applied
to individual radio--signal time series.
The Morlet mother wavelet, consisting of a complex sine wave modulated by a
Gaussian, is selected to search for radio signal variability, with
the non--dimensional frequency $\omega_0$ satisfying
the admissibility condition~\citep{1992AnRFM..24..395F}.
The WT is calculated for the period range from 10 to 600 seconds (or from 1 to
200\,seconds) with scales sampled as fractional powers of two with ${\delta j = 0.4}$, which is
small enough to give an adequate sampling in scale, a minimum scale
${s_{j}=9.64}$\,seconds (0.96\,second) and a number of scales $N=101$. 
Both the calculated
significance of the derived WT periodicities and the cone--of--influence 
are taken into account as described in \cite{2017SoPh..292....1K}. 
The value of the confidence level is set to 99\,\% for this article.
\\
\\
As a reference for the X--ray emission of the flare we select 
the {\it Fermi} mission {\it Gamma-ray Burst Monitor} ({\it Fermi}/GBM, 
\opencite{2009ApJ...702..791M}) data in the 26\,--\,50\,keV range.
Dedicated {\it Object Spectral Executive} (OSPEX) software from 
the {\tt spex} library of SolarSoft
\citep{1998SoPh..182..497F} is exploited for data preparation.
The same data set has been analyzed also in the article by
\cite{2016ApJ...830..101B}, where more instrumental details can be found.

\begin{figure}[t]
\centering
\includegraphics[width=12cm, bb = 15 12 555 380]{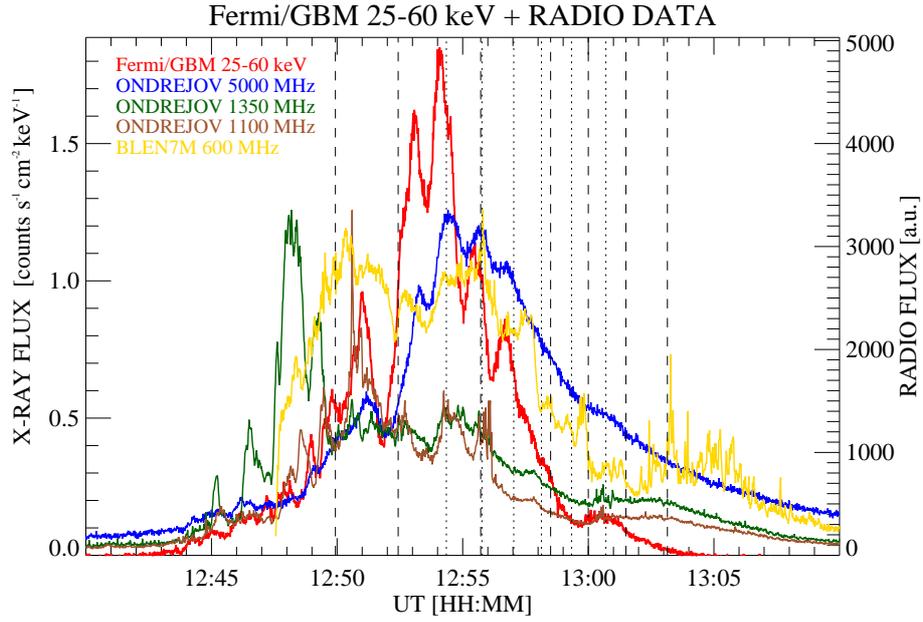}
\caption{Global overview of the 18 April 2014 solar flare: 
{\it Fermi}/GBM 26\,--\,50\,keV
light curve and the radio fluxes at 5000, 1350, 1100, and 600\,MHz.
The vertical dotted lines designate times of the EIS emission peaks
and the vertical dashed lines show times of the IRIS emission peaks
(for exact time and spatial location information on these peaks see 
Section\,3.}
\label{figure1}
\end{figure}

\begin{figure}
\centering
\includegraphics[height=6.5cm, bb =  4 7 404 414, clip=true]{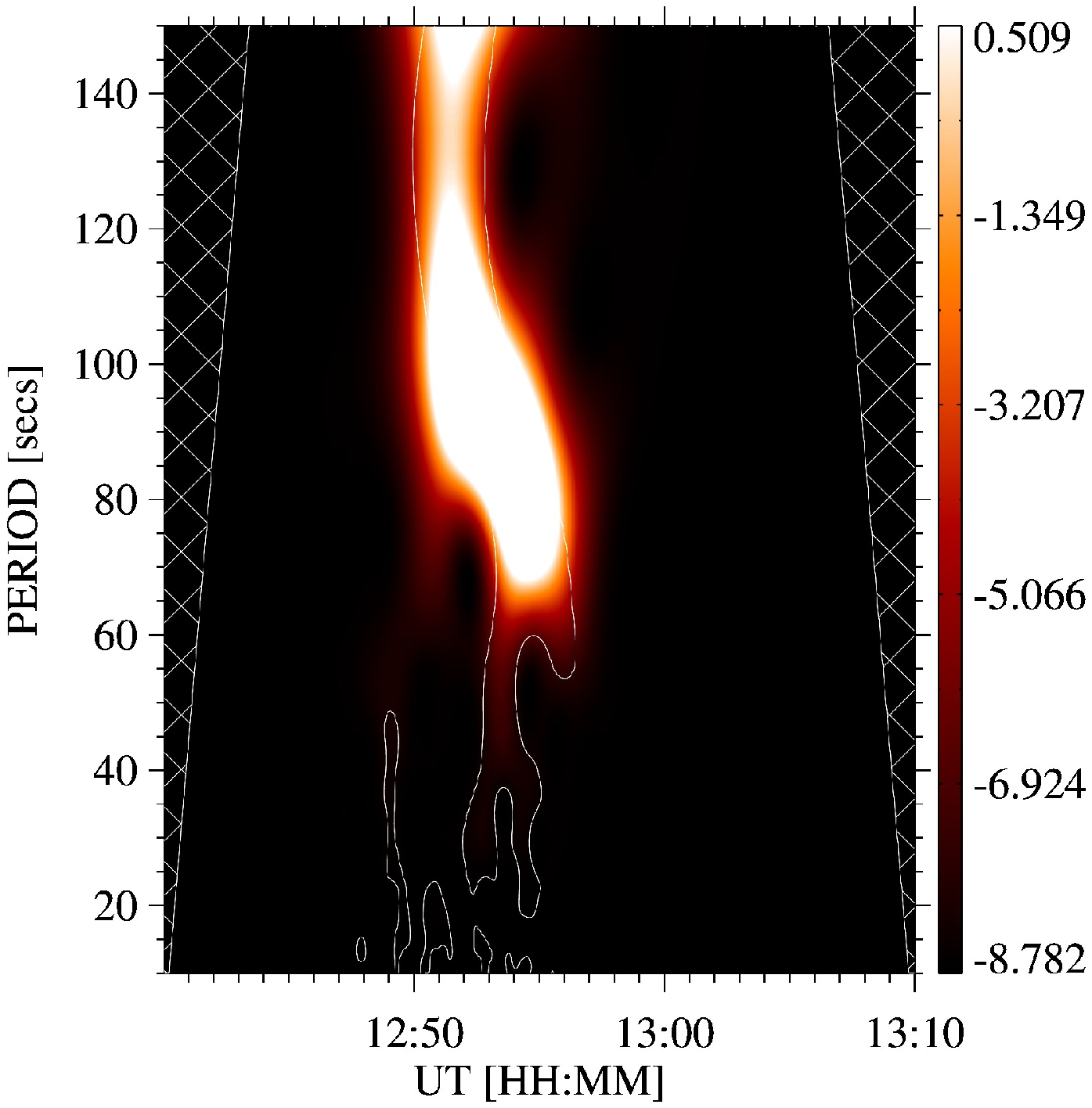}
\includegraphics[height=6.5cm, bb = 30 7 387 414, clip=true]{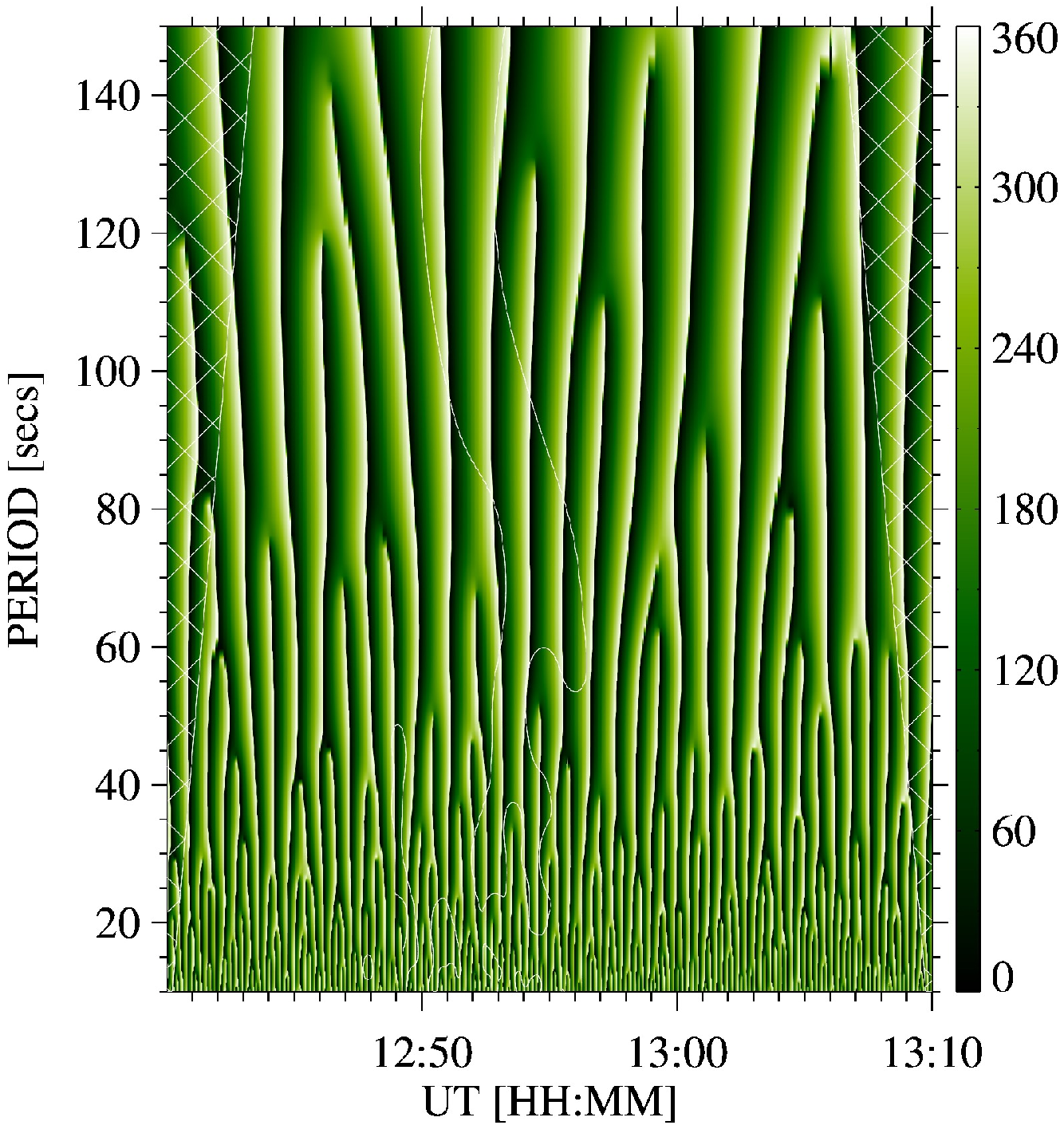}
\caption{Wavelet power and phase spectra of the {\it Fermi}/GBM 26\,--\,50\,keV X--ray emission.
The white contours mark the time--frequency domain area of the statistically 
significant periodicities while the cross--hatched areas show the 
cone--of--influence.}
\label{figure2}
\end{figure}

\section{Results}

A global overview of the 18 April 2014 solar flare is shown in
Figure\,\ref{figure1} using the {\it Fermi}/GBM 26\,--\,50\,keV light curve 
and the particular radio fluxes at 5000, 1350, 1100, and 600\,MHz. 
All these full--disk flux curves reveal a quasi--periodic behavior with
peaks which are more or less synchronized in general.

In Figure~\ref{figure1} the vertical dotted lines designate times of EIS emission
peaks (12:54:21, 12:55:46, 12:57:02, 12:58:08, 12:59:20, and 13:00:42 UT)
(\opencite{2016ApJ...830..101B}), and the vertical dashed lines show times of
IRIS emission peaks (12:49:56, 12:52:26, 12:55:43, 12:58:30, 13:00:00,
13:01:30, and 13:03:09 UT) (\opencite{2015ApJ...810...45B}). However, these
emission peaks are detected in the spatially very localized positions of the
slits of these instruments (EIS: $X=512\pm1$\arcsec, $Y=-219\pm2$\arcsec, IRIS:
$X=551\pm0.16$\arcsec, $Y=-186.4\pm0.16$\arcsec).

It is clearly seen that some of these EIS and IRIS peaks coincide  well
with the presented hard X--ray and radio fluxes peaks, for example, at 12:54:21 or
12:55:46\,UT.
On the other hand, in the time interval 12:57\,--\,13:00\,UT, where several EIS and
IRIS peaks are observed, the hard X--ray 26\,--\,50\,keV flux and the 
5000, 1350, and 1100\,MHz
radio flux plots are more or less smooth, except for the peaks present in the 600\,MHz
radio flux plot.

To know the characteristic periods in these processes, we analyze the hard
X--ray {\it Fermi}/GBM 26\,--\,50\,keV emission of this flare. The resulting wavelet
power and phase spectra are shown in Figure~\ref{figure2}. As seen here, the
flare in hard X--rays has in the time interval 12:49\,--\,12:57\,UT a
characteristic period at about 115\,seconds which displays shortening in the
course of
this time interval down to about 65\,seconds. This result, derived using the WT
method, is in coincidence with the previously derived value of 72$\pm$2.7\,seconds
(\opencite{2016ApJ...830..101B}) which was determined just from temporal
moments of the peaks in the interval 12:53\,--\,12:57\,UT. Our initial value of
115\,seconds reflects the appearance of the hard X--ray peak located at 12:51\,UT
which appears before the very regular next four peaks.

\begin{figure}[t]
\centering
\includegraphics[height=3.4cm, bb = 20 22 368 208, clip=true]{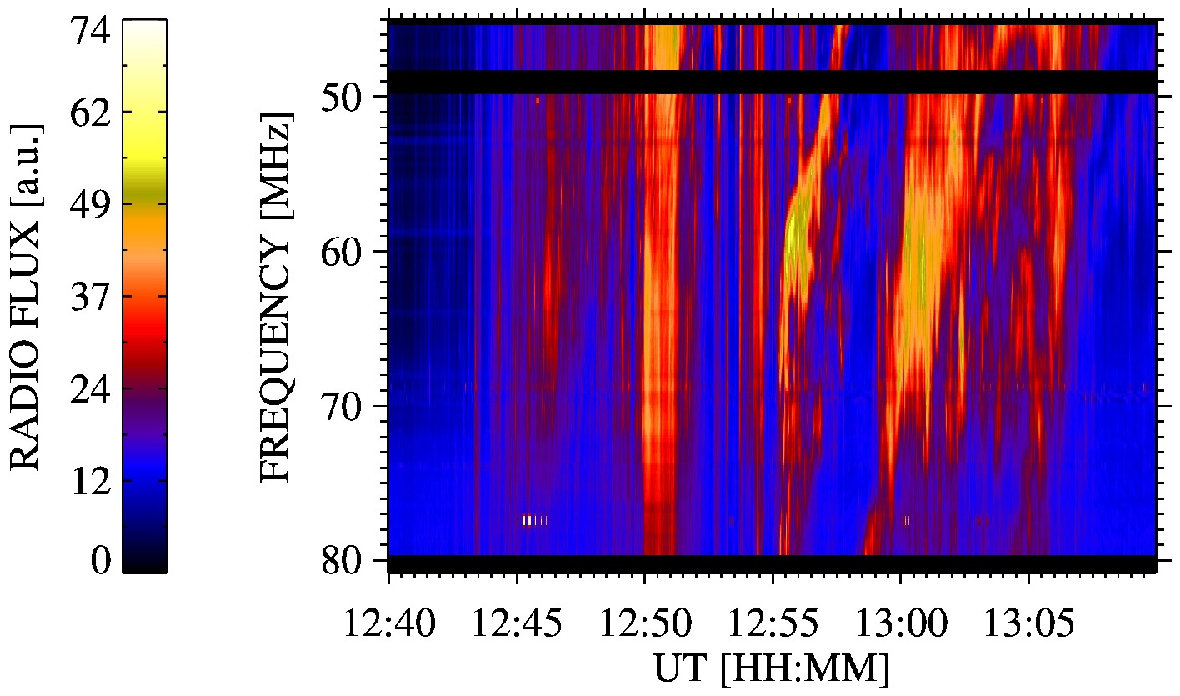}
\includegraphics[height=3.4cm, bb =114 22 415 208, clip=true]{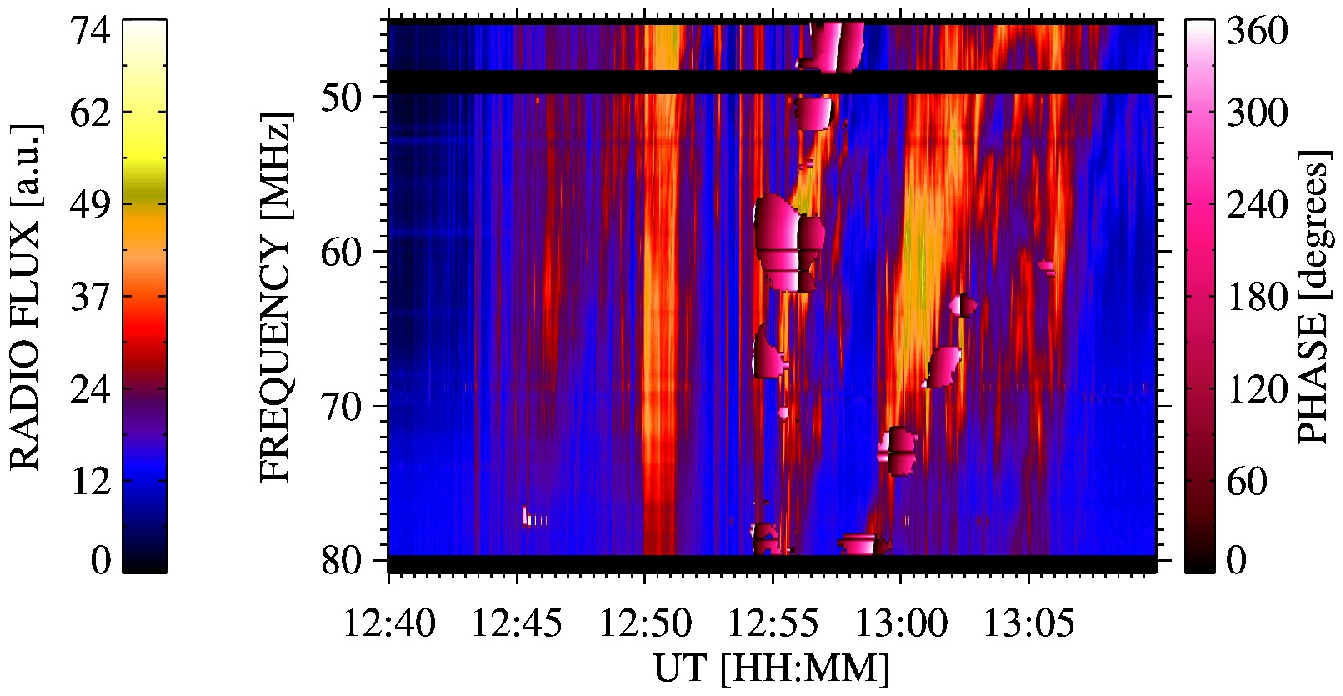}
\includegraphics[height=3.4cm, bb = 20 22 368 208, clip=true]{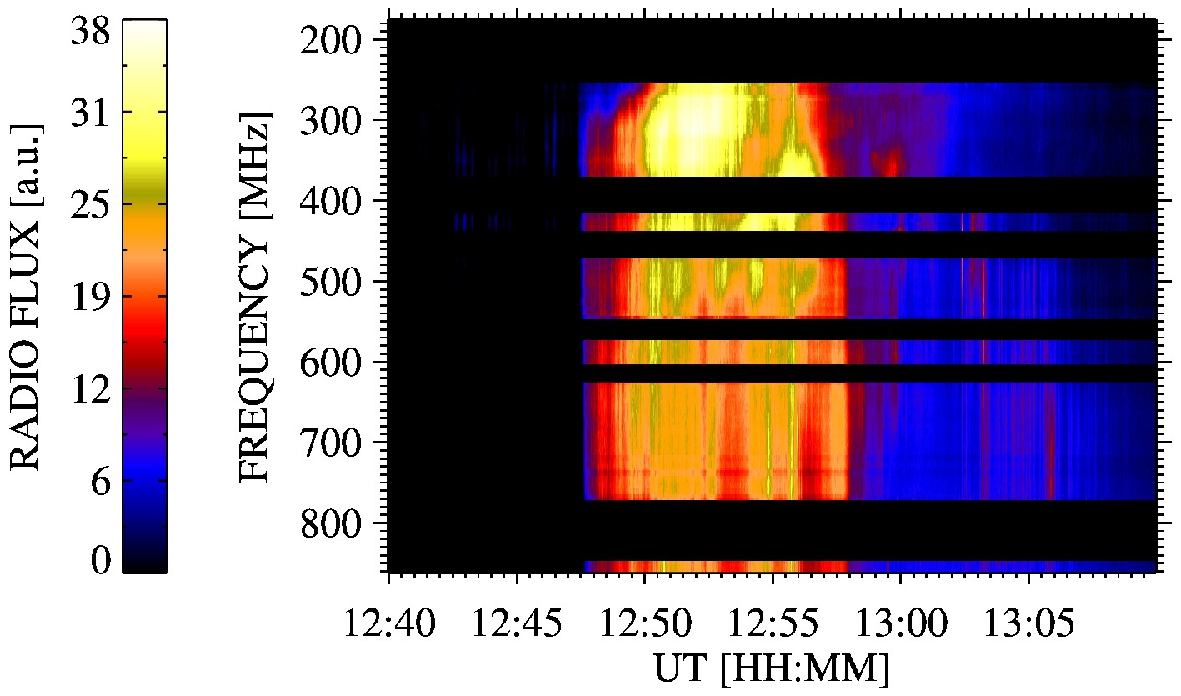}
\includegraphics[height=3.4cm, bb =111 22 415 208, clip=true]{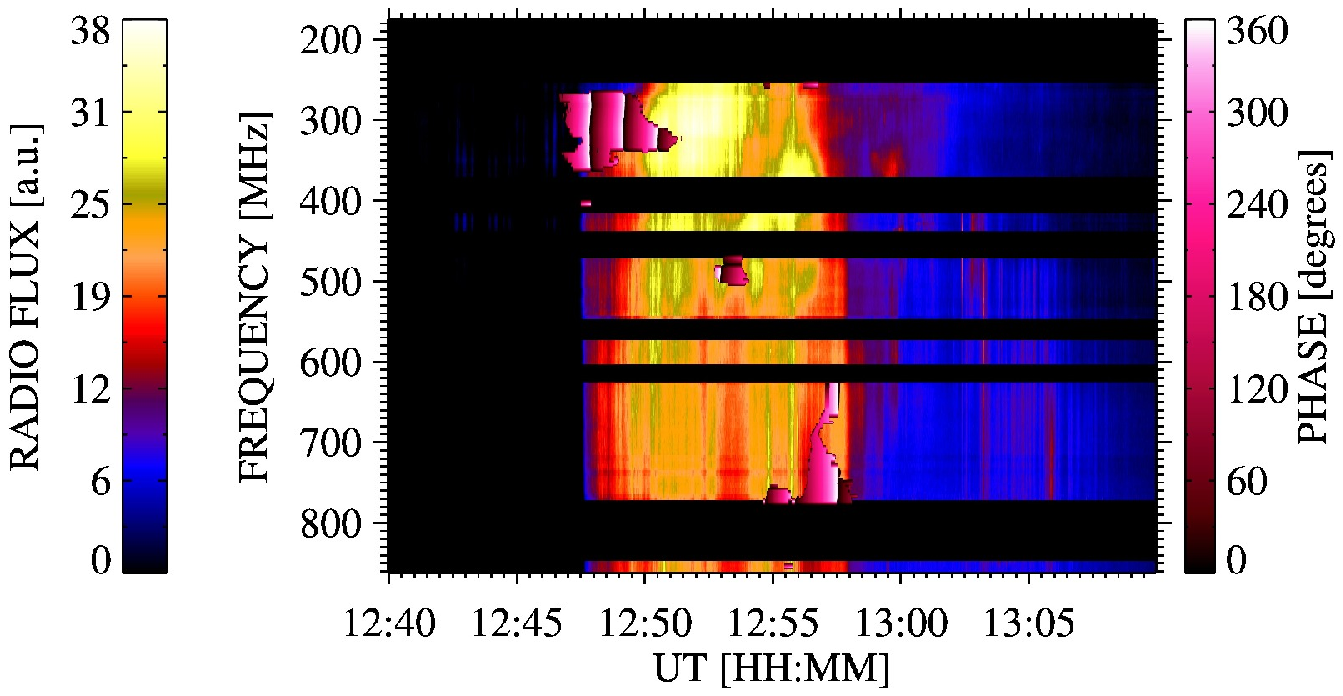}
\includegraphics[height=3.4cm, bb = 18 22 368 208, clip=true]{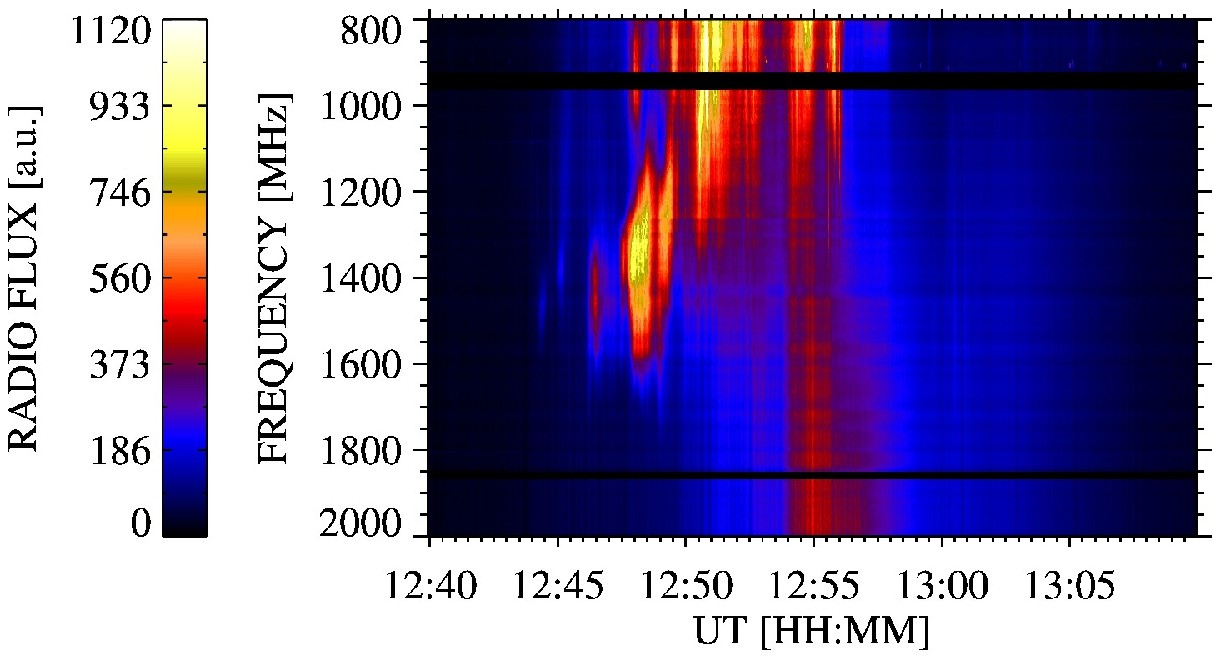}
\includegraphics[height=3.4cm, bb =108 22 415 208, clip=true]{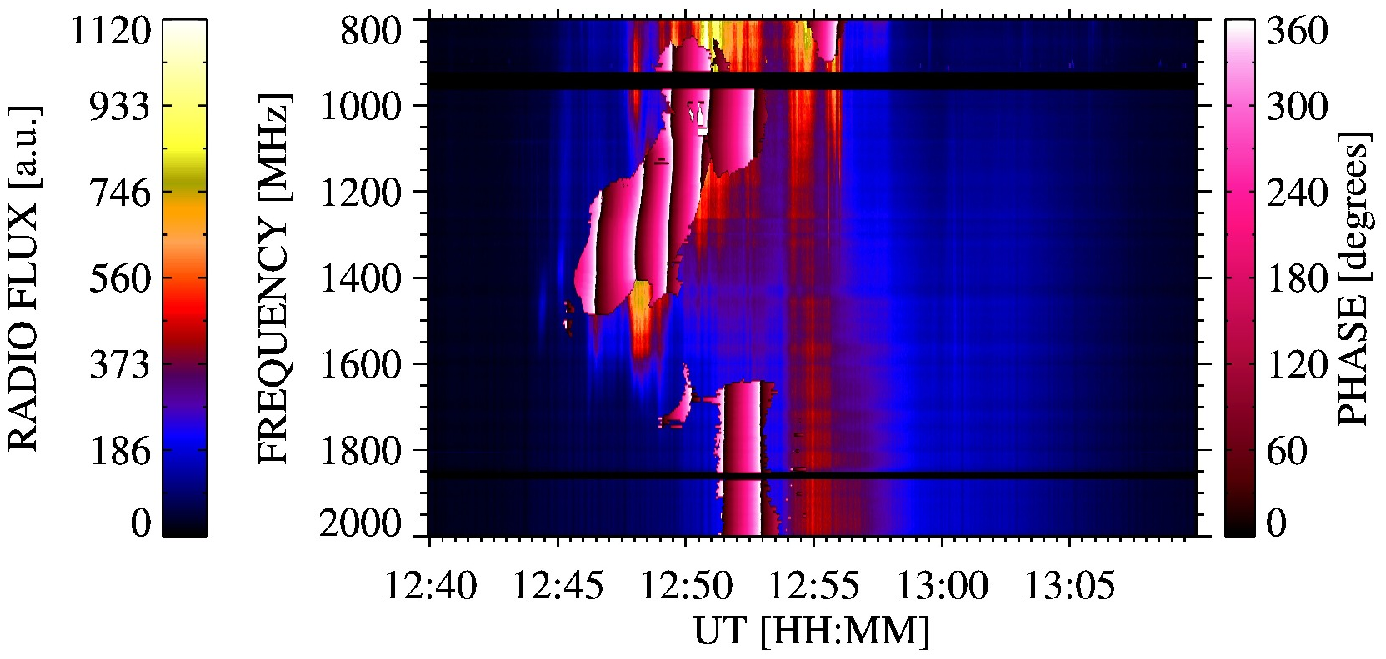}
\includegraphics[height=3.6cm, bb = 20 10 368 208, clip=true]{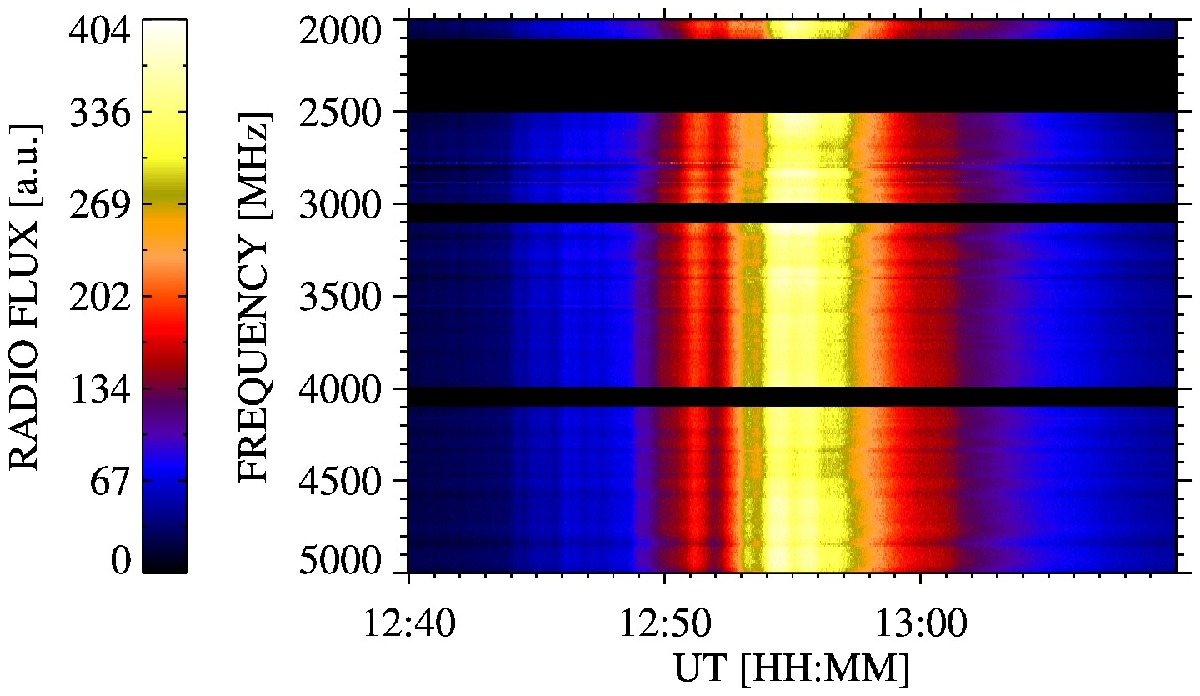}
\includegraphics[height=3.6cm, bb =108 10 415 208, clip=true]{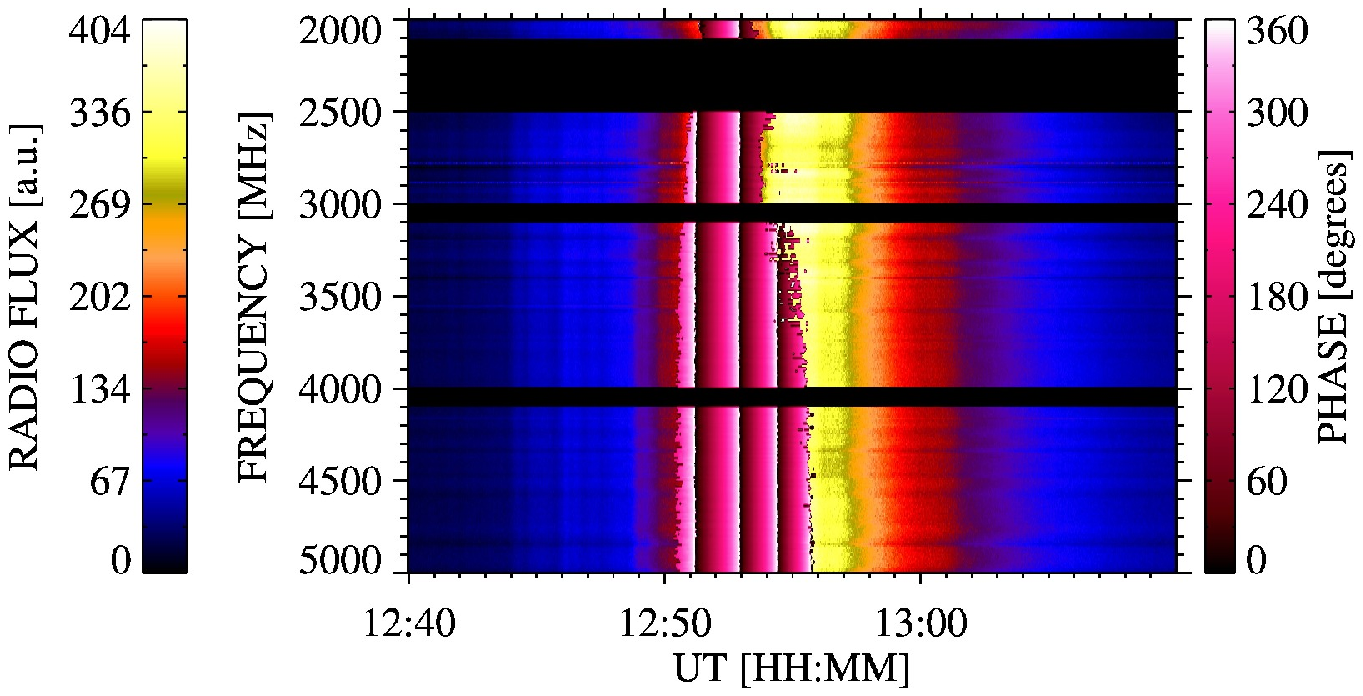}
\caption{Left column: the radio spectra observed by the e--Callisto radio spectrometers
(Glasgow: 45\,--\,80\,MHz, BLEN7M: 175\,--\,862\,MHz), and by the
Ond\v{r}ejov radiospectrographs (800\,--\,2000, and 2000\,--\,5000\,MHz).
Right column: the same radio spectra and the corresponding phase maps overplotted
on these radio spectra for periods detected within the 65\,--\,115\,second range.
The black lines in the phase pink bands correspond to zero phases.}
\label{figure3}
\end{figure}

Knowing the interval of characteristic periods in hard X--rays, we
construct the oscillations map for these periods for all radio spectra in
the 45\,--\,5000\,MHz frequency range. The radio spectra
(Figure~\ref{figure3}, left column) as well as the corresponding maps
of phase (pink bands overplotted on the radio spectra) are presented in
Figure~\ref{figure3} (right column). Note that lower frequencies in this figure
are presented above higher ones, which expresses that the emission at lower
frequencies is generated at higher heights in the solar atmosphere than that at
higher frequencies. Thus, the radio spectra and phase maps give us
information about flare processes and oscillations in the vertical direction in
the solar atmosphere.

As seen in the radio spectra, in a broad range of frequencies from 45 to
5000\,MHz (Figure~\ref{figure3}, left column), the flare appears in radio 
at about 12:44\,UT with a so-called drifting pulsation structure (DPS), lasting
till 12:53\,UT in the 800\,--\,1600\,MHz range and also appears as a series 
of type III bursts in the 45\,--\,80\,MHz range. At higher frequencies
(2000\,--\,5000\,MHz) there is a continuum that starts at about the same
time as the DPS and has its maximum at 12:54\,UT in coincidence with the hard
X--ray maximum. In the 175\,--\,862\,MHz range we see broadband emission
(12:47\,--\,13:08\,UT) consisting of fast--drifting type III bursts and
two slowly drifting bursts (at 12:49\,--\,12:55\,UT and
12:54\,--\,12:56 UT in the 500\,--\,250\,MHz range), designated by
\citet{2016ApJ...833...87C} as Continuum A and Continuum B, which look to be
the high--frequency precursors of the type II burst observed at
12:55\,--\,13:07\,UT in the 45\,--\,80\,MHz range.

In the right column of Figure~\ref{figure3}, there are the corresponding maps
of oscillations with periods in the selected 65\,--\,115\,seconds
interval. These oscillations are mostly at higher frequencies starting at about
12:45\,UT at 1400 MHz. In the 250\,--\,5000\,MHz range they are nearly
synchronized (within a few seconds). 
Only at the beginning of the radio flare in the interval 12:45\,--\,12:49\,UT and 
1200\,--\,1500\,MHz range we see some drift of the oscillation phase.
Namely, the phase drift in the bursts generated by the plasma emission
mechanism indicates a propagating wave with a period 65\,--\,115\,seconds. Details
about these phase drifts are given below. At lower frequencies, in the
45\,--\,80\,MHz range, the oscillations with a 65\,--\,115\,seconds period as a
part of a type II burst are detected later than those at higher
frequencies and sporadically only.

\begin{figure}[t]
\centering
\includegraphics[width=10cm, bb = 15 8 415 195, clip=true]{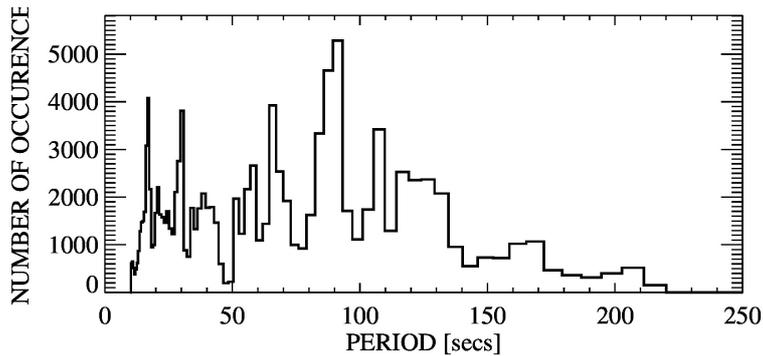}
\caption{The histogram of the periods in the 800\,--\,2000\,MHz radio spectrum
for the 12:44\,--\,12:54\,UT time interval.}
\label{figure4}
\end{figure}

\begin{figure}
\centering
\includegraphics[width=12.0cm,height=4.7cm, bb = 15 22 415 210, clip=true]{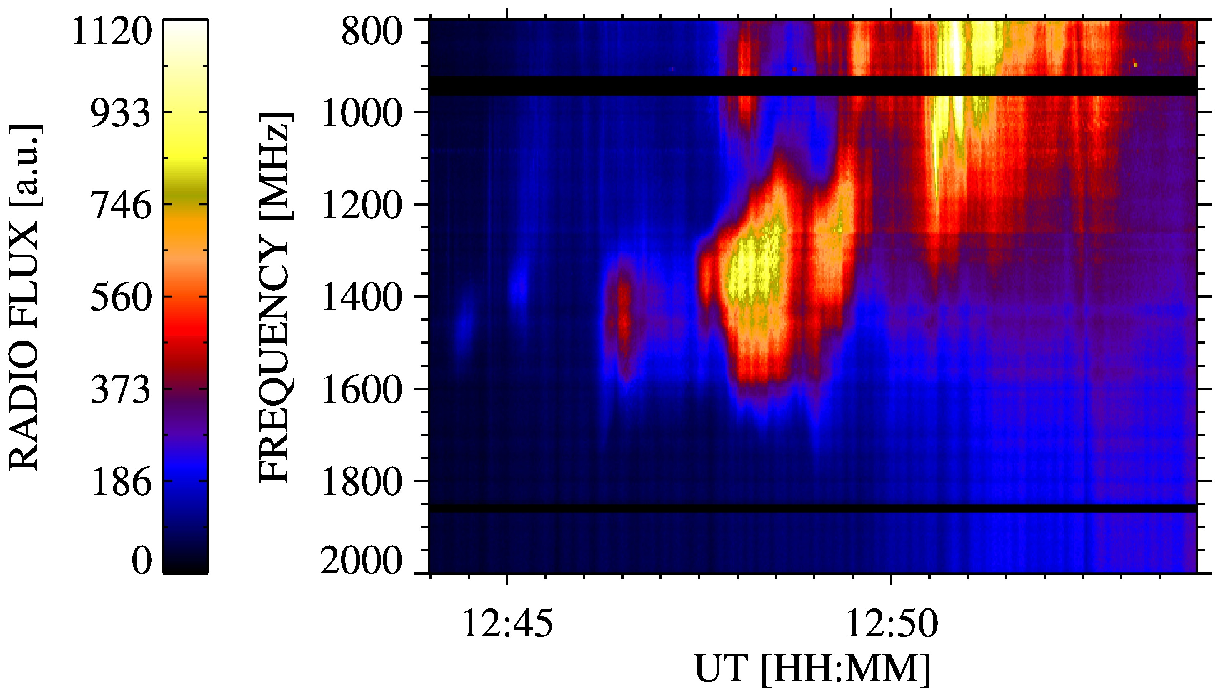}
\includegraphics[width=12.0cm,height=4.7cm, bb = 15 22 415 210, clip=true]{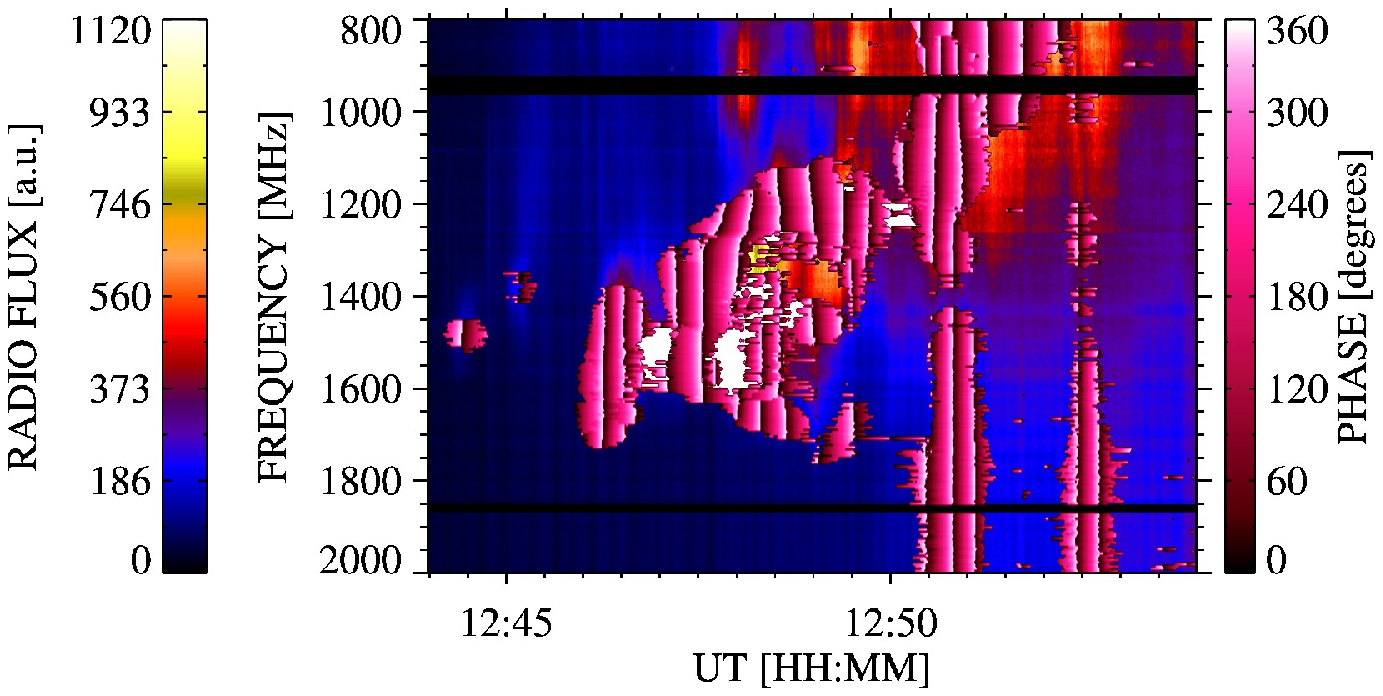}
\includegraphics[width=12.0cm,height=4.7cm, bb = 15 22 415 210, clip=true]{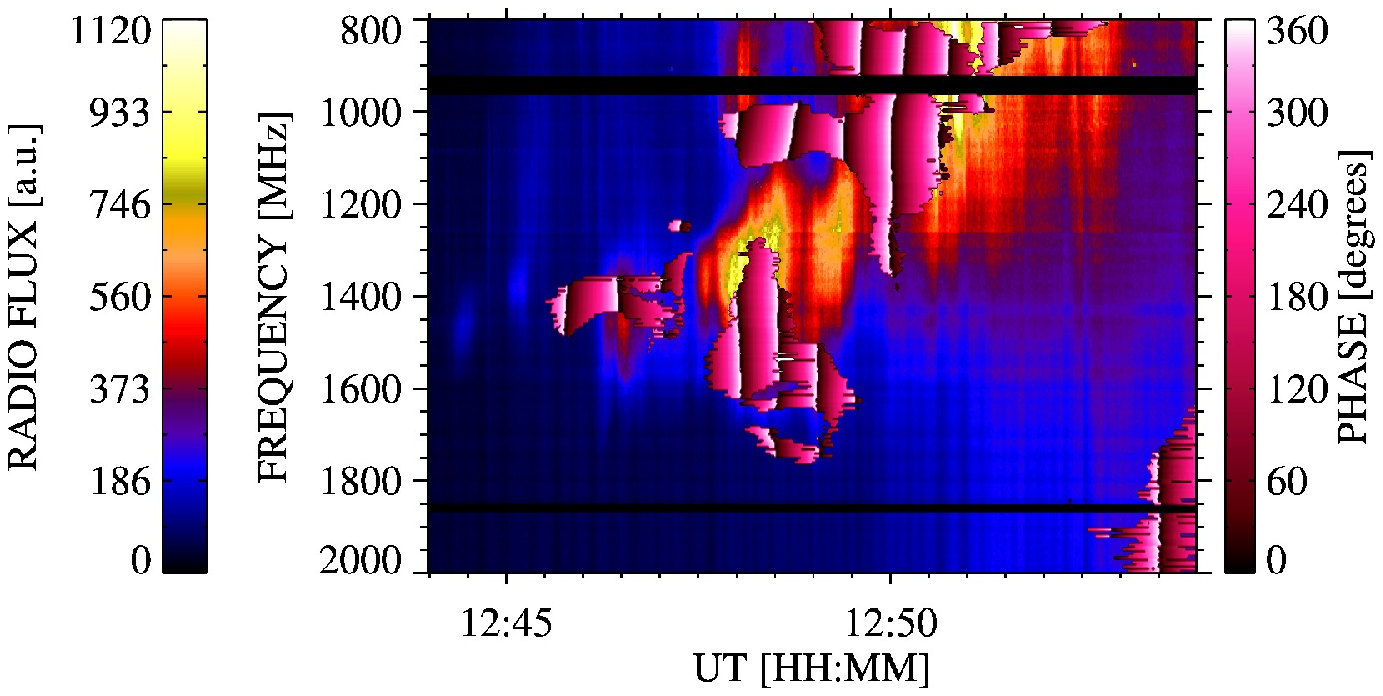}
\includegraphics[width=12.0cm,height=4.7cm, bb = 15  7 415 210, clip=true]{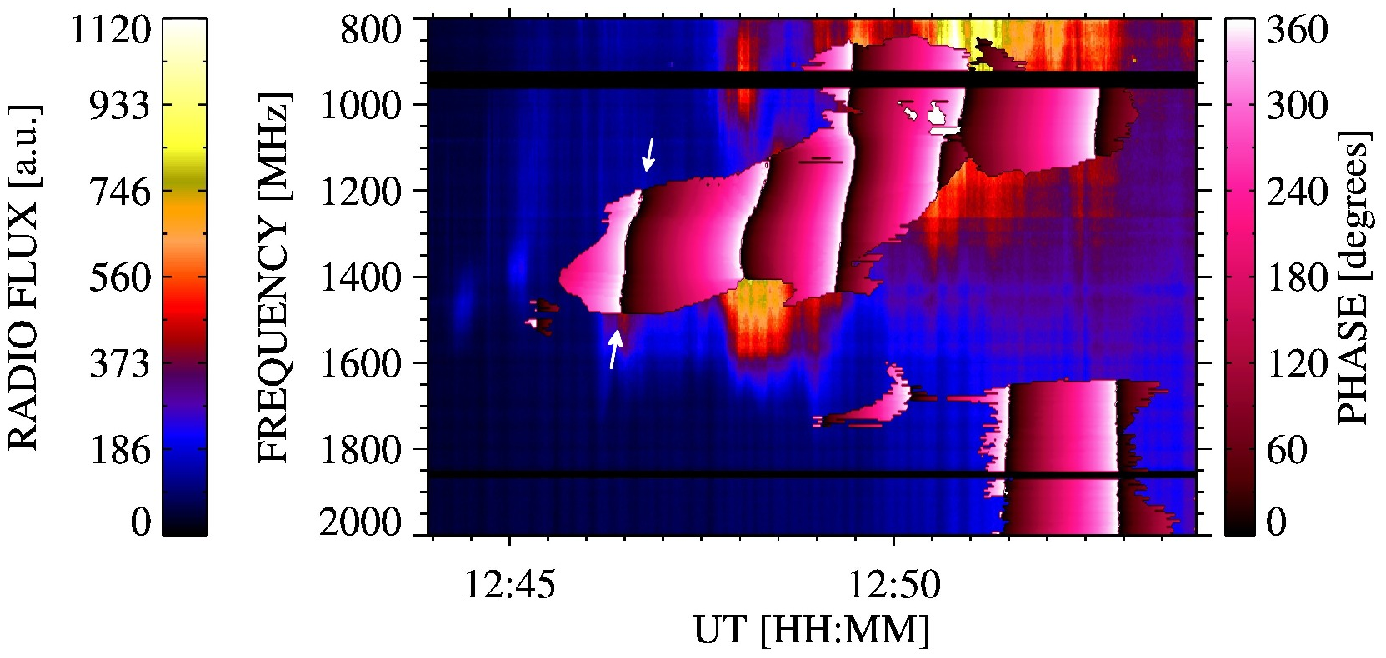}
\caption{The 800\,--\,2000\,MHz Ond\v{r}ejov radio spectrum in the 12:44\,--\,12:54\,UT time interval
(top panel) and the wavelet phase maps overplotted on this radio spectrum
for period ranges 10\,--\,30, 30\,--\,50, and 65\,--\,115\,seconds (bottom panels).
The white arrows in the bottom panel show the zero phase drifting from high to lower frequencies.}
\label{figure5}
\end{figure}

\subsection{Phase Maps of the Drifting Pulsation Structure Observed in the
800\,--\,2000\,MHz Range}

As mentioned above the flare radio emission stars in the DPS 
observation at 12:43\,UT and around 1400\,MHz. 
This relatively narrowband ($\approx$400\,MHz) emission consists
of many fast drifting pulses, and drifts as a whole from 1400\,MHz to 800\,MHz
(Figure\,\ref{figure3}, left column, third panel and detailed view in
Figure\,\ref{figure5}, top panel). Models of such emission were presented by
\opencite{2000A&A...360..715K}, \opencite{2004A&A...417..325K},
\opencite{2007A&A...464..735K}, \opencite{2008A&A...477..649B},
\opencite{2010A&A...514A..28K}, \opencite{2011ApJ...737...24B},
\opencite{2011ApJ...733..107K}, and \opencite{2015ApJ...799..126N}. It was
proposed that the DPS is the signature of a plasmoid formed in the early phase
of the fast magnetic reconnection in the flare current sheet. During the
plasmoid formation electrons are rapidly accelerated
(\opencite{2008ApJ...674.1211K}). A part of these super--thermal electrons are
trapped in the plasmoid, where they generate the plasma waves which are then
transformed into electromagnetic (radio) waves observed as the DPS. If the
injection of superthermal electrons into the plasmoid is quasi--periodic then
DPS is formed from separate pulses like the DPS presented by
\cite{2000A&A...360..715K}. On the other hand, if the injection is more or less
continuous then DPS looks like a narrowband drifting continuum with some pulses
(present case). Owing to the limited range of densities inside the plasmoid, the
emission frequencies are limited as well. Therefore the frequency range of the
DPS gives us information about the density range in the plasmoid. Using the
relation for the plasma density $n_{e} (\rm{cm^{-3}}) = {\it f}^2/8.1 \times
10^{-5}$, where $f$ is the observed frequency in MHz \citep{1965sra..book.....K},
we obtain $n_e$ = 1.8$\times$10$^{10}$ -- 3.2$\times$10$^{10}$ cm$^{-3}$ at
12:48\,UT.

The DPS slowly drifts to lower frequencies with a frequency drift
$-1.66$\,MHz\,s$^{-1}$. To estimate the velocity of the corresponding plasmoid
some model of the solar atmosphere has to be used. 
There are several such models as presented in the papers
by~\cite{2007SoPh..244..167P,2013A&A...557A.115K,2016ApJ...833...87C}. However,
for the frequencies of the present DPS (above 800 MHz and the fundamental emission
of DPS), only the Aschwanden model~\citep{2002SSRv..101....1A} can be used.
Therefore, in the following we use the Aschwanden model and its multiplied
version. We start with the Aschwanden model, but consider also that the
density in the plasmoid is higher than outside of the plasmoid. It can be shown
that in the case of the plasmoid observed in the 5 October 1992 flare
\citep{1998ApJ...499..934O}, the density inside the plasmoid was 6.9 times
higher than outside. If we assume for the present DPS (plasmoid) the same
density contrast as in the 5 October 1992 plasmoid, the derived upward
velocity of the plasmoid is 28 km s$^{-1}$. Now, let us utilize the result
of~\cite{2016ApJ...833...87C}. Namely, they determined the level of the second
harmonic frequency 445 MHz ($n_e$ = 6.11 $\times$ 10$^8$ cm$^{-3}$) in this
flare to be at the height 140 Mm. Multiplying the Aschwanden model by the factor
10.5, we obtain the same density at the same height. Then using this multiplied
Aschwanden model and the same procedure mentioned above, we determine the
upward velocity of the plasmoid to be 73 km s$^{-1}$.
(Note that according to the DPS model \citep{2008A&A...477..649B}
the velocity of the plasmoid is given by the tension of surrounding magnetic 
field lines. Thus, the plasmoid velocity can be in the interval from zero 
velocity (motionless case) up to the local Alfv\'en velocity (maximum velocity 
case). The estimated plasmoid velocities are realistic because they are lower 
than the Alfv\'en velocity expected in the magnetic reconnection site.)
The negative frequency
drift of the DPS shows that the plasmoid moves upward in the solar atmosphere.
Namely, during its upward motion the density as well as plasma frequency inside
the plasmoid decreases and thus the DPS drifts towards lower frequencies. The
fast--drifting pulses show individual injection of the super--thermal electrons
into the plasmoid. Note that other super--thermal electrons that are not
trapped in the plasmoid escape in the upward direction as the observed type III
bursts and downwards generating the observed continuum.

The DPSs are very important bursts in any solar flare and therefore we make a
detailed analysis of the  periodicities in a shorter time interval of
12:44\,--\,12:54\,UT  where DPS appears (Figure~\ref{figure5}, top
panel). We compute the histogram of the significant periods in this radio
spectrum (Figure~\ref{figure4}). As seen here, the radio spectrum consists of
periods in a broad range from seconds up to 200\,seconds, which agrees with the
idea of cascading (multi--scale and multi--periodic) processes in the flare
magnetic reconnection (\opencite{2011ApJ...733..107K} and
\opencite{2011ApJ...737...24B}).

The oscillation maps for this radiospectrogram for several intervals of periods
are shown in Figure~\ref{figure5} together with the radiospectrogram (top
panel). They confirm that the processes generating the DPS are multi--scale
and multi--periodic as the significant oscillations are of a co--temporal and
co--spatial appearance. An interesting aspect is that the zero phase (black
line) with the period interval 65\,--\,115\,seconds drifts from a frequency
of about 1500\,MHz to 1200\,MHz in the time interval from
12:46:30\,--\,12:46:40\,UT (Figure~\ref{figure5}, bottom part, see the white
arrows). 
The frequency drift is about $-30$\,MHz\,s$^{-1}$, which corresponds
to velocities of about 500\,km\,s$^{-1}$ and 1300\,km\,s$^{-1}$ in
the Aschwanden and the 10.5 $\times$ Aschwanden density models, respectively,
where we use also the correction to the enhanced density in the plasmoid.
We propose that this drift indicates the presence of a fast magnetoacoustic
wave propagating upwards in the solar atmosphere. Note that such a wave was
detected in the numerical simulations just after a sudden enhancement of the
resistivity at the start of the magnetic reconnection in the current sheet
\cite{1988BAICz..39...13K}. We think that this wave modulates the plasma
density in the radio source and thus modulates the radio emission as proposed in
the papers by \cite{2013A&A...550A...1K} and \cite{2013A&A...552A..90K}.

\begin{figure}[t]
\centering
\includegraphics[width=12.0cm,height=3.0cm, bb =  0 13 427 107, clip=true]{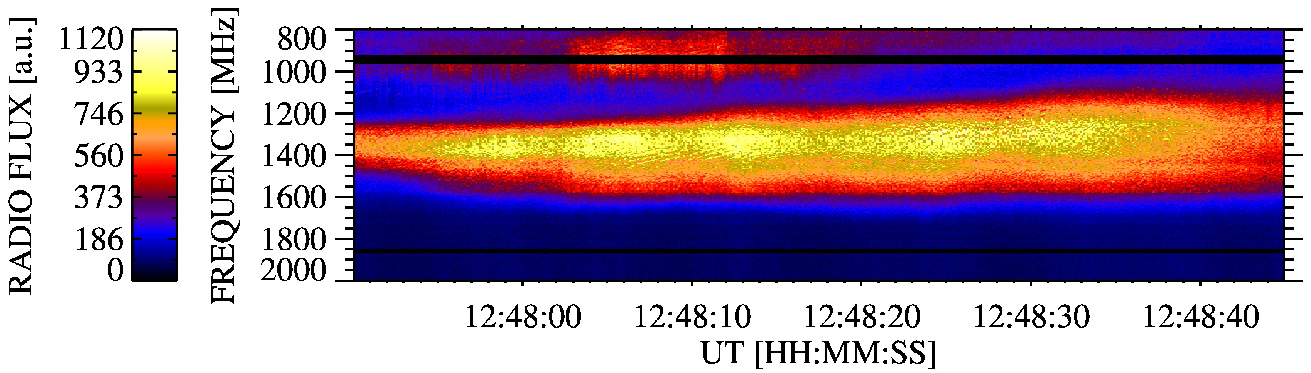}
\includegraphics[width=12.0cm,height=3.0cm, bb =  0 13 427 107, clip=true]{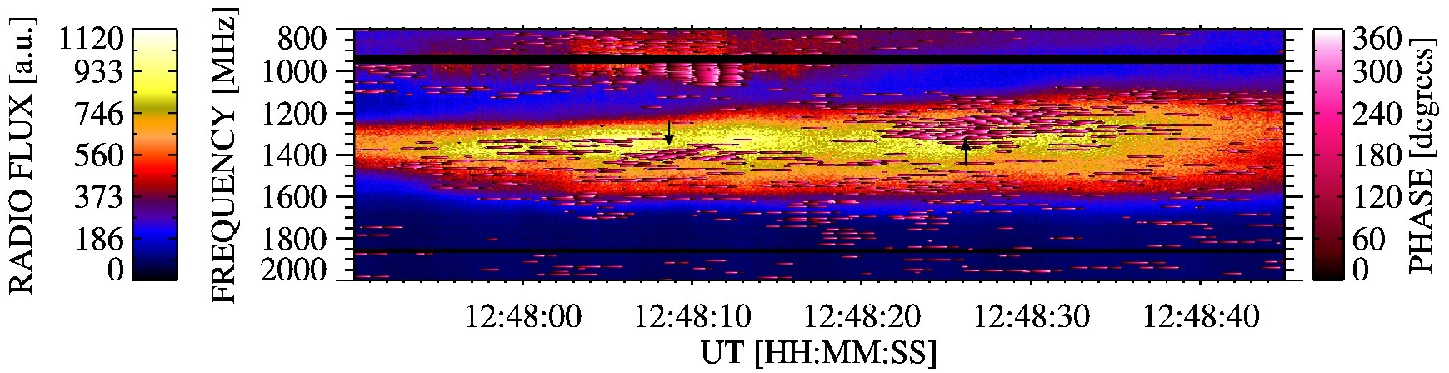}
\includegraphics[width=12.0cm,height=3.0cm, bb =  0  0 427 107, clip=true]{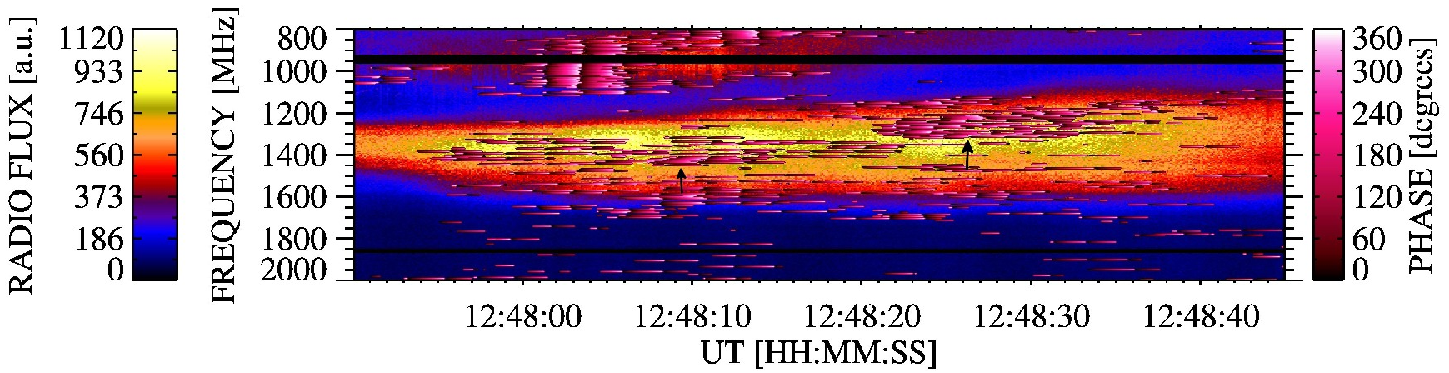}
\caption{The 800\,--\,2000\,MHz Ond\v{r}ejov radio spectrum in the 12:47:50\,--\,12:48:45\,UT time
interval (top panel) and the wavelet phase maps overplotted on this radio spectrum
for period ranges 1.1\,--\,1.7 and 1.8\,--\,3.0\,seconds (middle and bottom panels, respectively).
The black arrows in the middle and bottom panels show the regions with drifting phases.}
\label{figure6}
\end{figure}

Remarkable features detected in the 800\,--\,2000\,MHz frequency range only
during the temporal interval of just 120\,seconds (12:47\,--\,12:49\,UT)
stimulate us to analyze this radio spectrum using 0.1\,second data
temporal sampling (Figure~\ref{figure6}). Here, the periods 1.1\,--1.7\, and
1.8\,--\,3.0\,seconds are recognized not only in the DPS structure, but also
in the type III bursts in the 800\,--\,1100\,MHz range at
12:48:00\,--\,12:48:20\,UT. The phases overplotted on DPS, which start at
about 1400\,MHz and at 12:48:07\,UT (and last till 12:48:11\,UT) drift
towards lower frequencies with a frequency drift of about $-28$\,MHz\,s$^{-1}$.
Moreover, if we enlarge the phase map in the 1100\,--\,1300\,MHz
range at 12:48:21\,--\,12:48:32\,UT, we can see phases drifting negatively as
well as positively. A detailed analysis of the original radio spectrum
(Figure~\ref{figure6}, top panel) reveals that these drifting phases correspond
to very weak fiber bursts superimposed on the DPS structure. Their
frequency drift is of course the same as that of their phases 
($-28$\,MHz\,s$^{-1}$)
because they occurr on the radio spectrum repeatedly with a
characteristic period of about second.

Generally, there are three types of models of the fiber bursts. Models
interpreting their frequency drift based on whistler waves
\citep{1987SoPh..110..381M}, models interpreting this drift with Alfv\'en
waves \citep{1990A&A...236..242T}, and models considering the fast
magnetoacoustic waves \citep{2006SoPh..237..153K,2013A&A...550A...1K}. The
frequency drift of the 65\,--\,115\,s oscillation phase found at the beginning
of DPS at 12:46:30\--\,12:46:40\,UT (Figure~\ref{figure5}, bottom part)
($-30$\,MHz\,s$^{-1}$) is comparable to that of the fiber bursts superimposed on
DPS (Figure~\ref{figure6}) ($-28$\,MHz\,s$^{-1}$). Therefore, if both these
features are generated in the same region with similar plasma parameters and
the drifting phase at 12:46:30\--\,12:46:40\,UT is a signature of the fast
magnetoacoustic wave then the fiber bursts in the present case are in better
agreement with the fast magnetoacoustic-wave (period about 1 second) type
or Alfv\'en-wave type models of fibers than that based on the whistler waves. 
For details, see the papers by
\cite{1987SoPh..110..381M,1990A&A...236..242T,2013A&A...550A...1K}.

\subsection{The Phase Maps of the 175\,--\,862\,MHz Radio Spectrum in the Time 
Interval of EIS Oscillations}

\begin{figure}[!t]
\centering
\includegraphics[width=10cm,bb = 4 33 420 204, clip=true]{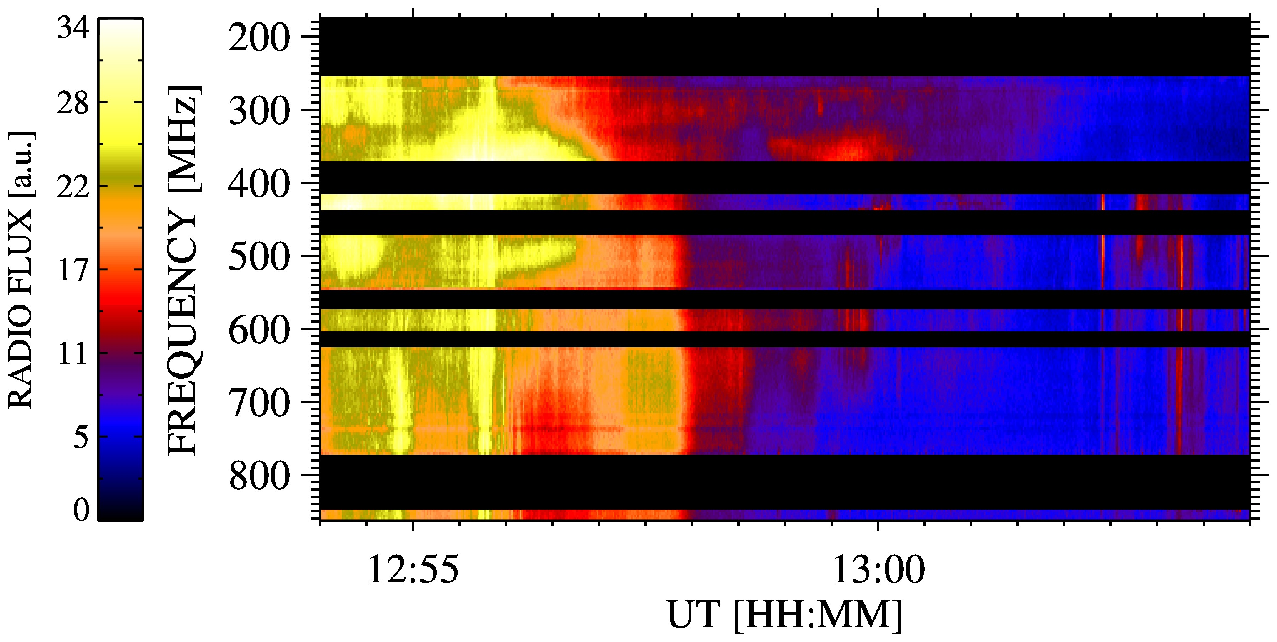}
\includegraphics[width=10cm,bb = 4 33 420 204, clip=true]{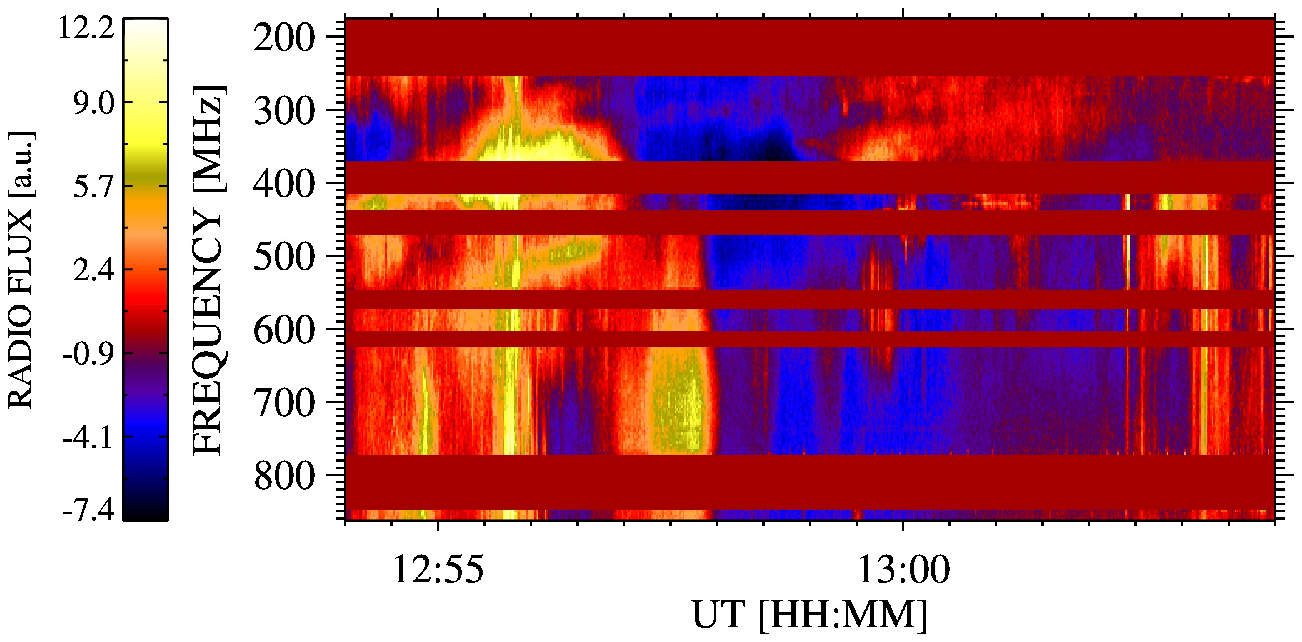}
\includegraphics[width=10cm,bb = 4 33 420 204, clip=true]{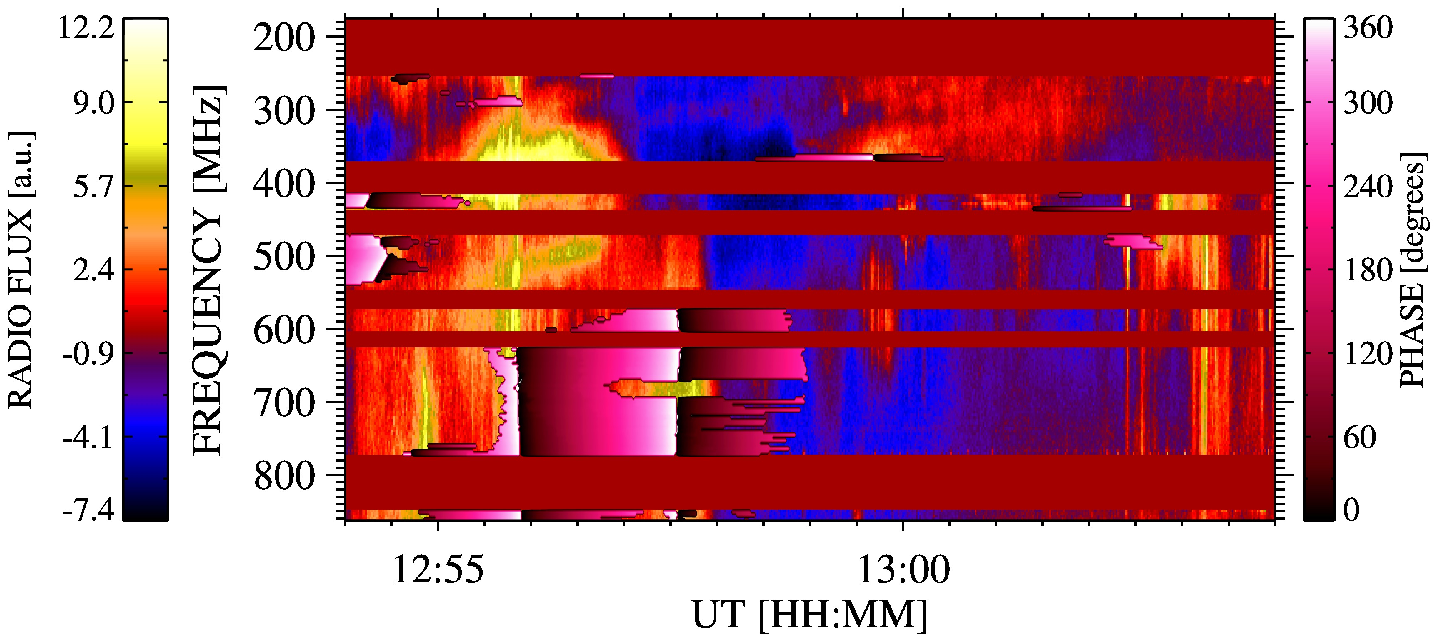}
\includegraphics[width=8.0cm,height=4.5cm,bb = -18 12 496 352, clip=true]{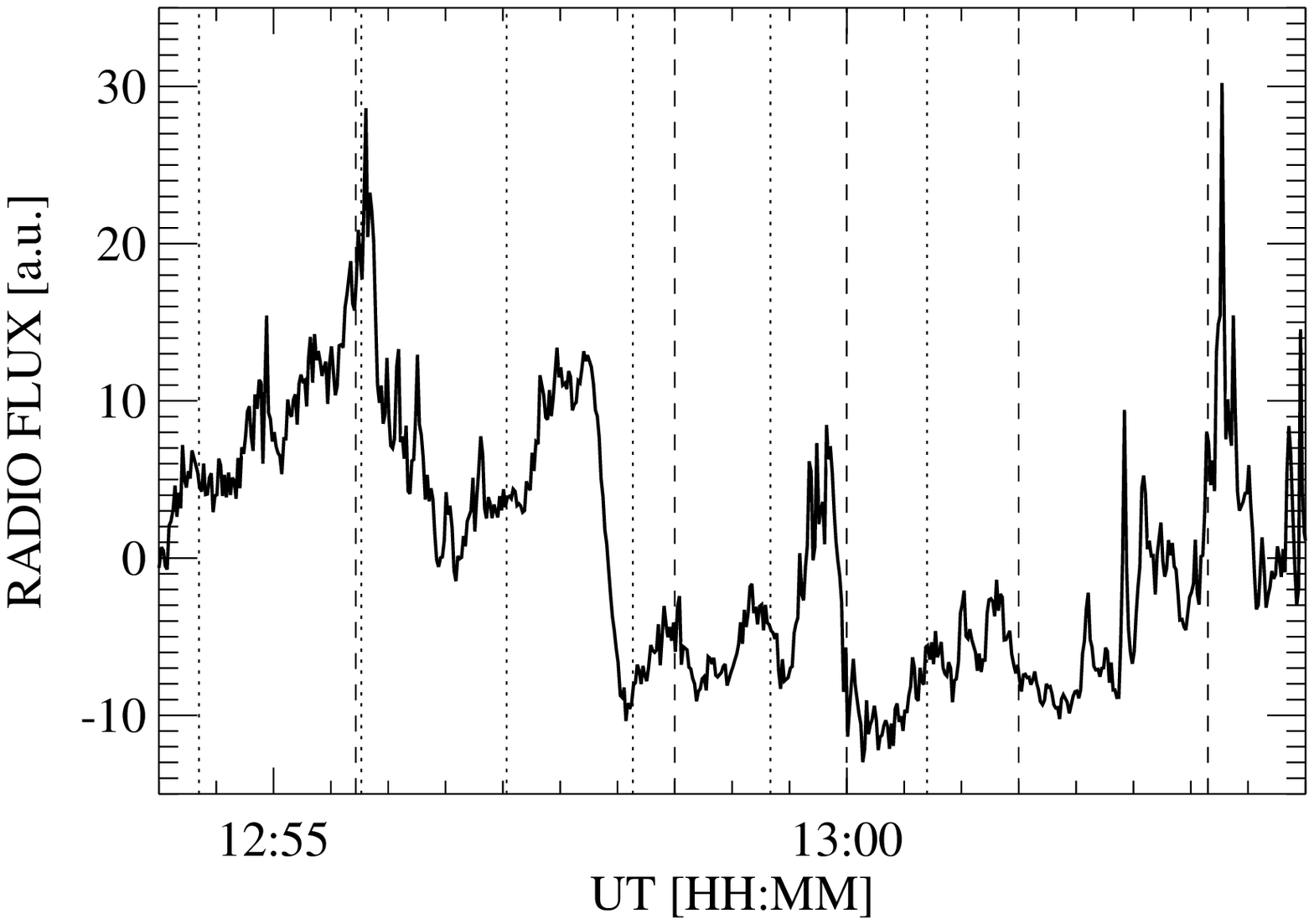}
\caption{The 175\,--\,860\,MHz Callisto Bleien radio spectrum in the 12:54\,--\,13:04\,UT time
interval:
top -- original spectrum; upper middle -- adapted spectrum (long-term trend removed);
bottom middle -- adapted spectrum overplotted by the phase map for the period 65\,--\,115\,s;
bottom -- an example of the adapted radio flux at the 600\,MHz frequency.
The vertical dotted and dashed lines in the 600\,MHz plot designate times of the
EIS and IRIS emission peaks, respectively (the same as in Figure~\ref{figure1}).}
\label{figure7}
\end{figure}

\cite{2016ApJ...830..101B} presented quasi--periodic 
({\it P}\,$\approx$\,75.6$\pm$9.2\,s)
intensity fluctuations in emission lines of O{\sc \,IV}, Mg{\sc \,VI}, 
Mg{\sc \,VII}, Si{\sc \,VII}, Fe{\sc \,XIV}, and Fe{\sc \,XVI} 
during the flare soft X--rays rise in the
12:54\,--\,13:01\,UT time interval (see also Figure~\ref{figure1} in the
present paper for the hard X--ray flux plot). They proposed that these
fluctuations are by a series of energy injections into the chromosphere by
non--thermal electron beams. Because in that paper there is no direct evidence
of these beams (the hard X--ray emission in this time interval is smooth, see
Figure~\ref{figure1}), we look for signatures of these beams in the
radio range. Seeing the plots of radio fluxes on several selected frequencies,
we find that only the 600\,MHz record shows some distinct variations during
the time interval of our interest. Therefore, we focus our attention on the
radio spectrum in the 175\,--\,862\,MHz range (Figure~\ref{figure7}, top part).
In this radio spectrum (at 12:55\,--\,13:00\,UT  in the 450\,--\,862\,MHz
range) we see groups of fast--drifting type III bursts. These bursts
are considered to be signatures of electron beams which thus supports the
explanation of emission line fluctuations proposed by
\cite{2016ApJ...830..101B}.

To  visualize the radio bursts even better, we form the new spectrum with long
periods (above 200\,s) removed. It is presented in Figure~\ref{figure7} (second
part from the top) and the corresponding 600\,MHz radio emission plot is
the bottom panel of this figure. However, as expressed by the vertical
lines in this 600\,MHz plot an association between the type III burst and the EIS
and IRIS peaks in some cases is good ({\it e.g.} at 12:55:46) and in other cases
problematic. This is probably due to radio spectrum being a whole--disk
record consisting of all bursts from any location, whereas the EIS and IRIS
peaks come only from the particular spatial locations of the slits of the EIS
and IRIS instruments.

We also check for a quasi--periodic character of the type III bursts in the
studied 12:54\,--\,13:54\,UT time interval. As shown in Figure~\ref{figure7}
(third panel from the top) only in limited intervals of time and
frequencies (12:55:40\,--\,12:57:30\,UT, 550\,--\,800\,MHz) do we find the
65\,--\,115\,seconds period oscillations, which is the same period found in the
{\it Fermi}/GBM 26\,--\,50\,keV light curve.

In the time interval from 12:49 to 12:56 UT two slowly drifting
features are observed in the 250\,--\,500\,MHz range (Continuum A and
Continuum B according to \citet{2016ApJ...833...87C}). As shown by
\citet{2016ApJ...833...87C} the Continuum A source was stationary and 
the source of Continuum B moved with the velocity 397 km s$^{-1}$.

\section{Conclusions}

We find that the 18 April 2014  M7.3 flare event as a whole has a 
quasi--periodic character in the 65\,--\,115\,seconds period interval.
This quasi--periodic behavior is recognized not only in the {\it Fermi}/GBM 
hard X--ray emission, but also in radio waves over a broad range of frequencies.
In most cases, similar to the 1 August 2010  C3.2 flare event
(\opencite{2017SoPh..292....1K}),
the phases of these 65\,--\,115\,seconds oscillations are nearly synchronized
(within a few seconds) over a very broad range of frequencies.
Therefore, similar to the paper by \cite{2017SoPh..292....1K}, we explain
this synchronization by fast electron beams (type III bursts) propagating
through the solar corona, with acceleration modulated by a wave or an
oscillation process with a period in the 65\,--\,115\,seconds interval.

However, at the very beginning of the flare at the time of the plasmoid
formation (expressed in radio as the DPS), in the phase map with the
65\,--\,115\, seconds period, we find a negative drift of the phase which
shows a propagating wave. We propose that this wave is the fast
magnetoacoustic wave propagating from the plasmoid site upwards. Such
a wave can be generated as follows: in the current sheet magnetic
reconnection starts when the current sheet is sufficiently narrow, 
{\it i.e.} when
the electric current density inside the current sheet is sufficiently high. If
this current density overcomes some threshold then the anomalous resistivity
can be suddenly generated \citep{1978A&A....68..145N}. Such a sudden
enhancement of the localized resistivity leads to a strong local heating and
increase of the pressure, which generates magnetoacoustic waves. Such a wave
was detected in the numerical simulations just after a sudden enhancement of
the resistivity at the start of magnetic reconnection in the current sheet
\citep{1988BAICz..39...13K}. In fragmented reconnection
\citep{2011ApJ...737...24B,2011ApJ...733..107K} there can be several such
regions with highly varying resistivity. They are located in space between
plasmoids. The generated magnetoacoustic wave then propagates over or out of
the plasmoids.

We think that this magnetoacoustic wave modulates the plasma density in the
radio source and thus modulates the radio emission as proposed in the papers by
\cite{2013A&A...550A...1K} and \cite{2013A&A...552A..90K}, and produces the
observed WT phase drift.

It is interesting that the frequency drift of this drifting phase ($-30$\,MHz
s$^{-1}$) (probably the fast magnetoacoustic wave) is comparable to that of
fiber bursts superimposed on DPS ($-28$\,MHz\,s$^{-1}$). In the present case, it
could speak in favour of the fast magnetoacoustic-wave type or Alfv\'en-wave
type models of fibers in comparison with the whistler-wave type model.

The drifting pulsation structure DPS reveals many periods ranging from 1 to
200\,s. It indicates that the DPS is connected with a multi--scale and
multi--periodic process. We even find fibers periodically repeated with a
$\approx$1\,second period and superimposed on the DPS. 

We also check for the periods found in the EIS and IRIS spatially localized
observations by \cite{2016ApJ...830..101B}. In the 450\,--\,862\,MHz range we
recognize the type III bursts (electron beams) as proposed by
\cite{2016ApJ...830..101B}, but their time coincidence with the EIS and IRIS
peaks is in some cases good and in others not so good. We think that this is due
to the radio spectrum beeing a whole--disk record consisting of all
bursts at any location. On the other hand, the EIS and IRIS peaks are emitted
only from the locations of the slits in the EIS and IRIS observations, 
and the flare process evolves in time as well as in space.

\begin{acks}
This research was supported by Grants 16--13277S and 17--16447S of the Grant
Agency of the Czech Republic.
This work was supported by the Science Grant Agency project VEGA 2/0004/16 (Slovakia).
Help of the Bilateral Mobility Project SAV--16--03 of the SAS and CAS is acknowledged.
This article was created by the realisation of the project ITMS No. 26220120029,
based on the supporting operational Research and development
program financed from the European Regional Development Fund.
The authors are indebted to the Institute for Astronomy, ETH Zurich, and FHNW Windisch
(Switzerland) and to the School of Physics and Astronomy of the University of Glasgow
(Scotland, UK) for the Callisto data.
This research has made use of NASA's Astrophysics Data System.
The wavelet analysis was performed with software based on tools provided by C.
Torrence and G. P. Compo at {\tt http://paos.colorado.edu/research/wavelets}.
\end{acks}

\section*{Disclosure of Potential Conflicts of Interest}
The authors declare that they have no conflicts of interest.

\bibliographystyle{spr-mp-sola}
\bibliography{ref_karlicky_rybak_monstein_v4}

\end{article}
\end{document}